\documentclass[twocolumn]{aastex6}

\usepackage{graphicx}
\usepackage{ulem}
\usepackage{amsmath}
\usepackage{wrapfig}
\usepackage{color,soul}

\fullcollaborationName{}

\begin{document}

\title{Accreting double white dwarf binaries: Implications for \textit{LISA}}

%% Authors with the same affiliation can be grouped in a single
%% \author and \affil call.
\author{Kyle Kremer, Katelyn Breivik}
\affil{Center for Interdisciplinary Exploration and Research in Astrophysics (CIERA), Department of Physics and Astronomy, Northwestern University \\
2145 Sheridan Rd, Evanston, IL 60201, USA}
\email{kremer@u.northwestern.edu, katelyn.breivik@northwestern.edu}

\and

\author{Shane L. Larson}
\affil{Center for Interdisciplinary Exploration and Research in Astrophysics (CIERA), Department of Physics and Astronomy, Northwestern University \\
2145 Sheridan Rd, Evanston, IL 60201, USA}
\affil{The Adler Planetarium \\
1300 S Lakeshore Drive, Chicago, IL 60605}
\email{s.larson@northwestern.edu}

\and

\author{Vassiliki Kalogera}
\affil{Center for Interdisciplinary Exploration and Research in Astrophysics (CIERA), Department of Physics and Astronomy, Northwestern University \\
2145 Sheridan Rd, Evanston, IL 60201, USA}
\email{vicky@northwestern.edu}

\begin{abstract}
We explore the long-term evolution of mass-transferring white dwarf binaries undergoing both direct-impact and disk accretion and explore implications of such systems to gravitational wave astronomy. We cover a broad range of initial component masses and show that these systems, the majority of which lie within the \textit{LISA} sensitivity range, exhibit prominent negative orbital frequency evolution (chirp) for a significant fraction of their lifetimes. Using a galactic population synthesis, we predict $\sim2700$ of these systems will be observable with a negative chirp of $0.1\ \rm{yr}^{-2} $ by a space-based gravitational-wave detector like \textit{LISA}. We also show that detections of mass-transferring double white dwarf systems by \textit{LISA} may provide astronomers with unique ways of probing the physics governing close compact object binaries.
\end{abstract}

\section{Introduction} \label{sec:intro}

Close binary systems containing compact objects are of relevance to a variety of fields in astronomy.  Compact binaries are a dominant source class for both ground and space-based interferometric gravitational wave detectors. Making up the most substantial fraction of close binaries are double white-dwarf (DWD) binaries \citep[e.g.,][]{Marsh1995}.

As gravitational radiation drives the components of a DWD binary closer together, it is possible for one of the stars to fill its Roche lobe, creating a semi-detached system. In the standard scenario, mass is pulled from the donor star through the system's inner Lagrangian point and enters into orbit around the accretor, eventually settling into an accretion disk \citep[e.g.,][]{Frank2002}. However, if the radius of the accretor star constitutes a large fraction of the binary separation (as is the case for DWD binaries), it is possible for the mass transfer stream to impact the accretor directly before a disk can form. The nature of this process, which is known as direct-impact accretion, determines whether the system remains stable (potentially forming an AM CVn \citep{Nather1981,Tutukov1996,Nelemans2004}) or unstable, leading to a merger and possibly a Type Ia supernova \citep[e.g.,][]{Webbink1984,Iben1984}.

For the case of disk accretion, tidal torques exerted between the binary orbit, the disc, and the component spins provide a mechanism for angular momentum to be transferred back into the orbit \citep[e.g.,][]{Soberman1997,Frank2002}. However, in the case of direct-impact accretion, such a mechanism does not obviously exist, making the long-term stability of systems which experience direct-impact accretion uncertain.

Several analyses have examined the evolution of DWD systems undergoing direct-impact accretion, including \citet{Marsh2004}, \citet{Gokhale2007}, \citet{Sepinsky2014} and, most recently, \citet{Kremer2015}. Each of these analyses concluded that, to varying degrees, a population of DWD binaries may emerge from a phase of direct-impact accretion as stable systems. The number of stable systems was shown in all analyses to depend sensitively upon the mass ratio of the system at the onset of mass-transfer and upon the strength of tidal coupling between the component stars and the binary orbit.

DWD systems are likely to be a dominant source of gravitational waves  (GWs) for space-based interferometers such as the Laser Interferometer Space Antenna (\textit{LISA};\citet{Amaro2013,Amaro2017}). For frequency intervals of interest, it is likely that DWD systems will define \textit{LISA}'s limiting confusion noise for frequencies below $\sim$ 3 mHz.\citep{Nelemans2001a,Nelemans2001b,Nelemans2001c,Liu2010, Ruiter2010,Yu2010}. Understanding the nature of the direct-impact mass-transfer process and its role in the evolution of DWD binaries undergoing gravitational radiation is therefore of prime importance to \textit{LISA}.

In this analysis, we explore the evolution of close DWD binaries undergoing both direct-impact and disk accretion in the context of \textit{LISA} using the methods presented in \citet{Kremer2015}. In Section \ref{sec:calculation}, we present the basic assumptions made in our analysis, review the methods of \citet{Kremer2015}, and present the equations necessary for the calculation of relevant GW parameters. In Section \ref{sec:results}, we analyze the results of our calculations and in Section \ref{sec:gxSimulation}, we present Galactic population synthesis models and in Section \ref{sec:discussion} discuss the implications of our results in the context of \textit{LISA}. We conclude in Section \ref{sec:conclusion}.

\section{Calculation of binary evolution} \label{sec:calculation}
\subsection{Basic assumptions}
Following \citet{Kremer2015} and \citet{Sepinsky2014}, we assume binaries with two white dwarfs (WDs) of (spherically symmetric) masses $M_A$ and $M_D$; volume equivalent radii $R_A$ and $R_D$ (given by the zero-temperature mass-radius relation of \citet{Verbunt1988}); and uniform rotation rates which are parallel to the orbits of the binaries. As discussed in \citet{Kremer2015}, we assume the binaries to remain in circular Keplerian orbits throughout their entire evolution. The initial semi-major axis is chosen such that the volume-equivalent radius of the donor is equal to the volume-equivalent radius of its Roche lobe as fit by \citet{Eggleton1983}.

As in \citet{Kremer2015}, we  assume the donors are Helium (He) WDs, which have masses $\leq 0.5$ $M_{\odot}$ \citep{Marsh2011}.

\subsection{Review of calculations of previous analyses}
To calculate the long-term evolution of DWD binaries, we follow the method of \citet{Kremer2015}, which is discussed at length in section 2 of that analysis, and can be summarized as follows. The orbital parameters of the DWD systems, which include the semi-major axis, $a$, eccentricity, $e$ (assumed to be zero throughout), component masses, $M_A$ and $M_D$, total mass $M = M_A + M_D$, and component rotation rates, $F_A$ and $F_D$,\footnote{In \citet{Kremer2015}, the donor and accretor rotation rates are referred to as $f_A$ and $f_D$. Here use use $F_A$ and $F_D$ to avoid confusion with the gravitational wave frequency, which we denote with $f$.} evolve due to the effects of three separate components: mass transfer, tidal forces, and gravitational radiation.

The effect of mass transfer on the orbital parameters is calculated using the ballistic mass-transfer calculations of \citet{Sepinsky2007}, which determine the changes to orbital parameters after a single mass-transfer event by integrating the three-body system consisting of the two stars and a discrete particle representing the mass transfer stream. See \citet{Sepinsky2007} and section 2.2 of \citet{Kremer2015} for more detail.

The effect of tidal forces upon the binary orbital angular momentum, $J$, is expressed as:
\begin{equation}
\label{Jorbdot}
\dot J_{\rm{orb, tides}} = \frac{k_AM_AR_A^{2}}{\tau_A}\omega_A + 
\frac{k_DM_DR_D^{2}}{\tau_D}\omega_D,
\end{equation}
where $\omega_i$ is the difference between the component and orbital spins and $\tau_A$ and $\tau_D$ are synchronization time-scales which determine the strength of tidal forces. \citet{Kremer2015} examines the cases of weak tides ($\tau_A=\tau_D=10^{15}$ years at contact) and strong tides ($\tau_A = \tau_D = 10$ years at contact) and noted that stronger tides lead to a greater number of stable systems relative to weak tides; see section 2.7 of \citet{Kremer2015} as well as \citet{Marsh2004} and \citet{Gokhale2007}. Here, we limit our analysis to the case of the 10-year synchronization timescale (strong tides) as this is considered the more realistic scenario for DWD binaries at present \citep{Fuller2014}.

From equation (\ref{Jorbdot}) and the equation for spin angular momentum of the individual components, $J_{\rm{spin}} = kMR^2\Omega$, the effect of tidal forces upon the orbital parameters is obtained. See section 2.3.2 of \citet{Kremer2015} for further detail.

Finally, the effect of gravitational radiation on the evolution of the binaries is given by:
\begin{equation}
 \label{adot-GR}
 \dot a_{\rm{GR}} = - \frac{64}{5}\frac{G^3}{c^5}\frac{M_AM_DM}{a^3}.
 \end{equation}

\subsection{Mass transfer and calculation of the Roche lobe}
The mass transfer rate is proportional to the difference between the radius of the donor and its Roche lobe. As discussed in \citet{Sepinsky2007}, for asynchronous and/or eccentric binaries, the shape and volume of the Roche lobe is dependent upon the eccentricity, true anomaly, $\nu$, and the donor's rotation rate, which is expressed in that analysis through the parameter: 
\begin{equation}
\label{scriptA}
\mathcal{A}_i(e,f,\nu) = \frac{{F_i}^2(1+e)^4}{(1+e \cos \nu)^3},
\end{equation}
with $i\in\{A,D\}$ for the accretor and donor, respectively.

Equations (47)-(52) of \citet{Sepinsky2007} show the dependence of the Roche lobe upon $\mathcal{A}$ and the mass ratio, $q$, in various regimes. In this analysis, we assume the eccentricity of the systems to remain zero, but allow the rotation rate of the donor to vary. As discussed in \citet{Kremer2015}, the dependence of the Roche lobe upon the donor's rotation can have a profound effect upon the stability of these systems as they undergo mass transfer.

Using the method outlined here, the long-term evolution of DWD binaries undergoing direct-impact accretion can be calculated from Roche-lobe overflow to one of two possible end results: the onset of unstable mass transfer or the formation of an accretion disc.

Recent numerical simulations \citep{Motl2007} of direct-impact accretion in DWD binaries have shown that mass transfer becomes dynamically unstable for $\dot{M}\sim 1-10\, M_{\odot}\,\rm{yr}^{-1}$. However, \citet{Motl2007} and other analyses  \citep[e.g.,][]{Han1999} also note that the onset of super-Eddington accretion ($\dot{M}_{\rm{Edd}} \sim 10^{-5} - 10^{-4}$; see section \ref{sec:Mdot_Edd}) may lead to the formation of a common envelope and merger.

In line with previous analyses \citep{Marsh2004,Kremer2015}, we adopt a value of $\dot{M} = 0.01\,M_{\odot}\,\rm{yr}^{-1}$ as our limit for dynamically stable mass transfer. However, given the results of the aforementioned studies, one may raise the question of whether the choice of $0.01\,M_{\odot}\rm{yr}^{-1}$ should be raised to higher values closer to $\sim 1\,M_{\odot}\,\rm{yr}^{-1}$ or lowered to the Eddington limit. It turns out the answer to this question is not important for the results of this study regarding the number of \textit{LISA}-detectable systems. As can be seen by comparing Figure \ref{fig:plot} with Figure \ref{fig:gxExample} in section \ref{sec:gxSimulation}, systems which exceed the Eddington limit in our calculations constitute less than a few percent of our \textit{LISA}-detectable systems.

\subsection{Disk accretion}
\label{disc}

\citet{Kremer2015} modeled systems exclusively through the direct-impact accretion phase, assuming that once a system becomes disk accreting, the systems will remain stable. This assumption was made on the basis that once a system enters a phase of disk accretion, it will remain in that phase or possible become detached as the orbital separation will grow due to the ongoing mass transfer \citet[e.g.][]{Frank2002}.

Here we expand upon the previous analysis by continuing to model the systems through phases of disk accretion. We model disk accretion using a treatment similar to that of \citet{Marsh2004}, which is outlined in what follows.

As discussed in \citet{Kremer2015}, the evolution of the semi-major axis can be separated into three components,
\begin{equation}
\label{adot}
\dot a = \dot a_{\rm{MT}} + \dot a_{\rm{tides}} + \dot a_{\rm{GR}},
\end{equation}
where $\dot a_{\rm{MT}}$, $\dot a_{\rm{tides}}$, and $\dot a_{\rm{GR}}$ are the effects of mass transfer, tidal forces, and gravitational wave radiation, respectively, upon the evolution of the semi-major axis. $\dot a_{\rm{GR}}$ is given by equation (\ref{adot-GR}) and $\dot a_{\rm{tides}}$ is given by equation (17) of \citet{Kremer2015}.

As shown in equation (6) of \citet{Marsh2004}, the change in the system's semi-major axis due to mass transfer (in the form of disk accretion) is given by:
\begin{equation}
\label{adotMT}
\frac{\dot{a}_{\rm{MT}}}{2a} = -\left[ 1 - q - \sqrt{(1+q)R_A} \right] \frac{\dot{M}_D}{M_D},
\end{equation}
where $q=M_D/M_A$ is the system's mass ratio and $\dot{M_D}$ is the mass-transfer rate of the system, which is calculated using equation (35) of \citet{Kremer2015}. It follows that the evolution of the masses of the two stars is given by:
\begin {equation}
\label{mdonor}
\frac{dM_D}{dt} = \dot{M}_D
\end{equation}
and
\begin{equation}
\label{maccretor}
\frac{dM_A}{dt} = -\dot{M}_D.
\end{equation}

Disk accretion is known to synchronize the component spins with the orbit on time-scales shorter than the circularization time-scale \citep[e.g.,][]{Zahn1977}. Therefore, we assume the donor and accretor both become instantaneously synchronized upon the onset of disk accretion and remain synchronized throughout. In the event that the system becomes detached and mass-transfer ceases, the donor and accretor spins are allowed to vary (in a self-consistent manner) until mass transfer resumes.

As in \citet{Kremer2015}, we assume these systems remain circular throughout their evolution, since they are born circular and the evolution is dominated by tides. Therefore,
\begin{equation}
\label{eccentricity}
\dot{e} = 0.
\end{equation}

Together, the equations (\ref{adot}), (\ref{mdonor}), (\ref{maccretor}), and (\ref{eccentricity}) can be integrated over time to calculate the long-term evolution of systems through disk accretion.

\subsection{Super-Eddington accretion}
\label{sec:Mdot_Edd}
As systems undergo both direct-impact and disk accretion, it is possible for the mass-transfer rate to exceed the Eddington limit, but remain below the instability limit (defined as 0.01 $M_{\odot} \rm{yr}^{-1}$). As in \citet{Kremer2015} and \citet{Marsh2004}, the Eddington accretion rate is calculated using a modified version of the calculation used by \citet{Han1999}:

\begin{equation}
\label{eq-43}
\dot M_{\rm{Edd}} = \frac{8\pi G m_p c M_A}{\sigma_T (\phi_{L1} - \phi_a -\frac{1}{2}\mathbf{v_i}^2 + \frac{1}{2}(\mathbf{v_i - v_{\omega}})^2)}
\end{equation}
where $\sigma_T$ is the Thomson cross-section of the electron, $m_p$ is the mass of a proton, $\mathbf{v_i}$ is the impact velocity of the accreted particle, and $\mathbf{v_{\omega}}$ is the spin-velocity of the accretor's surface at the point of impact, both measured in the co-rotating frame of reference. In our calculations, this expression yields $\dot{M}_{\rm{Edd}}$ values in the range of $10^{-5}-10^{-4}\,M_{\odot}\,\rm{yr}^{-1}$.

\citet{Kremer2015} took note of systems that experienced super-Eddington accretion, but did not change the mass-transfer physics during the super-Eddington phase. Along the lines of \citet{Han1999,Nelemans2001b,Marsh2004}, it was suggested that the consequence of sustained super-Eddington would be a merger, the same result as a dynamically unstable system.

Here, we expand on previous analyses by modifying the equations governing the mass-transfer process during phases of super-Eddington accretion. During super-Eddington accretion, we modify the mass transfer rate onto the accretor ($\dot{M}_A$; as calculated by equation (35) of \citet{Kremer2015}) as follows:

\begin{equation}
\dot{M}_{A} = 
\begin{cases}
	\dot{M}_{\rm{Edd}}, & \text{if } \dot{M}_D\geq \dot{M}_{\rm{Edd}}\\
	\dot{M}_{D}, & \text{otherwise}
\end{cases}
\end{equation}
In the case that $\dot{M}_A\geq \dot{M}_{\rm{Edd}}$, the mass transfer ceases to be conservative (i.e. $\dot{M}_A \neq \dot{M}_D$), and the excess mass transfer rate determines how much mass is ejected from the system:
\begin{equation}
\dot{M}_{\rm{eject}} = \dot{M}_{A} - \dot{M}_{\rm{Edd}}.
\end{equation}

\bigskip

Using the method outlined above, the long-term evolution of DWD systems can be calculated from the onset of Roche-lobe overflow through phases of direct-impact and/or disk accretion.

\subsection{Calculation of total chirp}
For binaries in circular orbits, the gravitational wave frequency, $f$, and the orbital frequency, $f_{\rm{orb}}$, are related by:
\begin{equation}
\label{forb-f}
f = 2 f_{\rm{orb}}.
\end{equation}
$f_{\rm{orb}}$ is related to the semi-major axis by
\begin{equation}
\label{forb-a}
f_{\rm{orb}} = \left( \frac{GM}{4\pi^2a^3} \right)^{1/2},
\end{equation}
where $M=M_A+M_D$.

It follows from equations (\ref{forb-f}) and (\ref{forb-a}) that:
\begin{equation}
\label{f-a}
f = \left( \frac{GM}{\pi^2a^3} \right)^{1/2}.
\end{equation}
Differentiation of equation (\ref{f-a}) gives the chirp:
\begin{equation}
\label{chirp}
\dot f = -\frac{3}{2}\left( \frac{GM}{\pi^2a^5} \right)^{1/2}\dot a.
\end{equation}

It follows from equations (\ref{chirp}) and (\ref{adot}) that the chirp can be written as
\begin{equation}
\dot f = -\frac{3}{2}\left( \frac{GM}{\pi^2a^5} \right)^{1/2} (\dot a_{\rm{MT}} + \dot a_{\rm{tides}} + \dot a_{\rm{GR}}),
\end{equation}
or more succinctly:
\begin{equation}
\label{fdot}
\dot f = \dot f_{\rm{GR}} + \dot f_{\rm{MT}}  + \dot f_{\rm{tides}}.
\end{equation}

By calculating each of these three components over the course of a binary's evolution we can determine the relative effects of gravitational radiation, mass transfer, and tidal forces on the chirp.

\section{Results} \label{sec:results}

\begin{figure}[b!]
\plotone{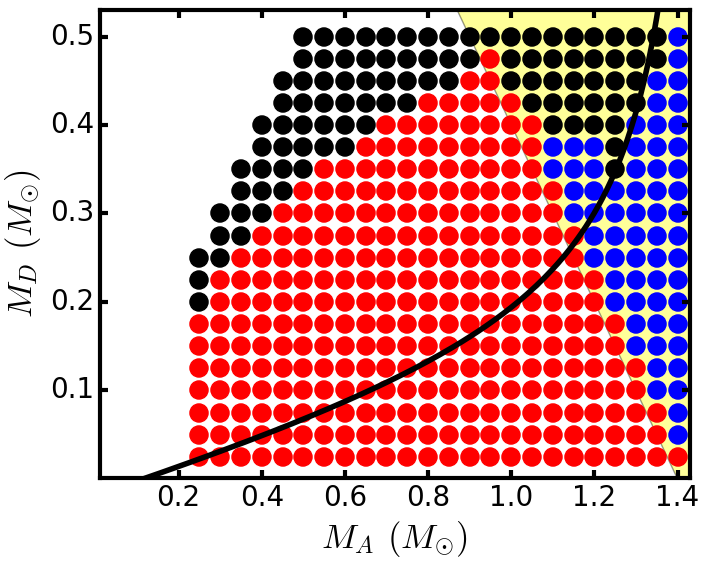}
\caption{\label{fig:plot} \footnotesize Results of long-term evolution of DWD systems of various initial masses. Red systems are stable throughout their lifetimes, through stages of both direct-impact and disk accretion. Black systems are unstable, meaning at some point they experience a mass transfer rate in excess of 0.01 $M_{\odot}\rm{yr}^{-1}$ at some point during their evolution. Blue systems  have an accretor which exceeds the Chandrasekhar limit during their evolution. The solid black line marks the boundary between disk and direct-impact accretion for initially synchronous and circular binaries. The region highlighted in yellow indicates systems with a total mass in excess of the Chandrasekhar limit. These systems are possible Type Ia supernova progenitors.}
\end{figure}

Figure \ref{fig:plot} shows the long-term evolution of DWD systems as calculated using the methods discussed in section \ref{sec:calculation}. 

Systems are evolved until the mass-transfer rate becomes unstable or until they reach a maximum evolution time of 10 Gyr. In Figure \ref{fig:plot}, systems marked by red circles evolve all the way to 10 Gyr without experiencing unstable mass transfer and are therefore classified as stable systems. Systems marked by a black circle are unstable systems, meaning they experience a mass transfer rate in excess of 0.01 $M_{\odot}\rm{yr}^{-1}$ at some point during their evolution and are expected to merge. Blue circle indicate systems whose evolution is stopped when the accretor reaches the Chandrasekhar limit (1.44 $M_{\odot}$).\footnote{Note that actually reaching the Chandrasehkar limit requires significantly more complex mass-transfer physics than that considered here. This is meant to serve as a basic approximation.} The region highlighted in yellow marks systems with a total mass in excess of the Chandrasekhar limit. Systems in this yellow region are considered possible Type Ia supernova progenitors.

The solid black line in Figure \ref{fig:plot} shows the boundary between disk and direct-impact accretion for initially synchronous and circular binaries (see \citet{Sepinsky2014} for more detail). All systems to the right of this boundary begin evolution in the disk accretion phase, and therefore never experience direct-impact accretion. Systems to the left of this boundary begin as direct-impact accretors, and eventually enter a disk phase. The time a system remains in the direct-impact phase varies from system to system, as discussed in \citet{Kremer2015}.

\begin{figure}[t!]
\plotone{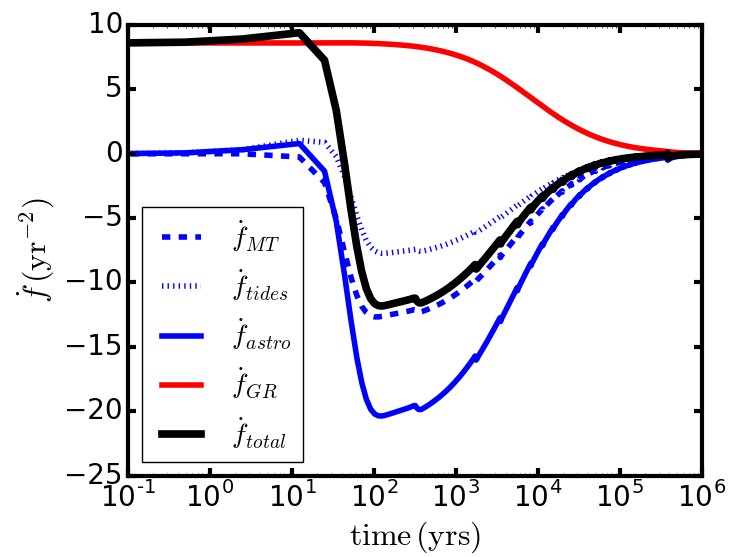}
\caption{\label{fig:chirp} \footnotesize Evolution of the different components of $\dot{f}$ during direct-impact accretion for a system with initial donor and accretor masses of 0.25 and 0.55 $M_{\odot}$, respectively. Here, $\dot{f}_{astro}$ is defined as the sum of $\dot{f}_{MT}$ and $\dot{f}_{MT}$. }
\end{figure}

Figure \ref{fig:chirp} shows the evolution of $\dot f_{\rm{MT}}$ (dashed blue), $\dot f_{\rm{tides}}$ (dotted blue), and $\dot f_{\rm{GR}}$ (solid red) for the first $10^6$ years of evolution of a system with initial donor and accretor masses of 0.25 $M_{\odot}$ and 0.55 $M_{\odot}$, respectively. Also shown is $\dot f_{\rm{astro}}$ (defined as sum of $\dot f_{\rm{MT}}$ and $\dot f_{\rm{tides}}$; solid blue) and the total chirp,  $\dot f$ (solid black), hereafter referred to as $\dot f_{\rm{total}}$.

\begin{figure}[t!]
\plotone{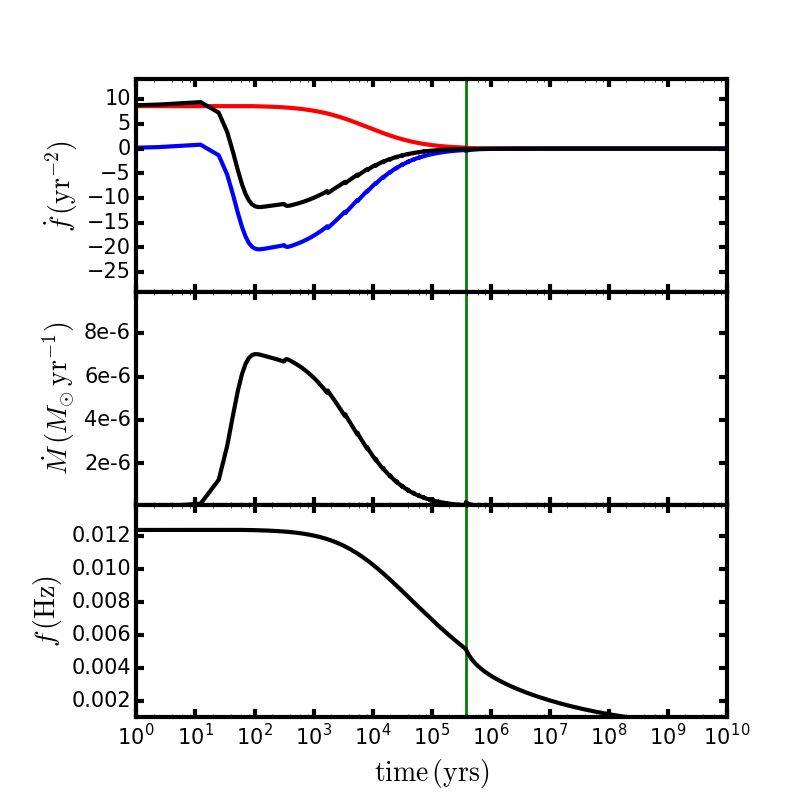}
\caption{\label{fig:evplot} \footnotesize Long-term evolution of the different components of $\dot{f}$ (top panel), the mass-transfer rate, $\dot{M}$ (middle panel), and the gravitational wave frequency of the system (bottom panel) for a system of initial donor and accretor masses of 0.25 and 0.55 $M_{\odot}$. The vertical solid green line marks the time of transition from direct-impact accretion to disk accretion, which for the system occurs at $3.79\times 10^5$ years.}
\end{figure}

Upon the onset of Roche-lobe overflow, mass transfer tends to drive the system to lower frequencies (larger orbital separation) in competition with gravitational radiation, which tends to drive the system to higher frequencies (smaller orbital separation), as is reflected by the negative and positive signs of $\dot f_{\rm{astro}}$ and $\dot f_{\rm{GR}}$, respectively, in Figure \ref{fig:chirp}. The relative strengths of these two components determine the overall sign of $\dot f_{\rm{total}}$. As Figure \ref{fig:chirp} shows, $\dot f_{\rm{total}}$ is dominated by $\dot f_{\rm{GR}}$ until mass transfer begins at which point $\dot f_{\rm{astro}}$ dominates. As tidal forces restore equilibrium between the component spins and the orbit, and mass transfer drives the components apart, all three components of $\dot f_{\rm{total}}$ approach zero. The mass transfer rate continues to decrease throughout the evolution, but never vanishes completely, driving the systems to ever-decreasing orbital frequencies, as shown in Figure \ref{fig:evplot}.

As discussed in detail in \citet{Kremer2015}, the length of the synchronization time-scale on which tidal forces act has a significant effect upon the systems' ability to adapt to mass transfer between the components. For the case of strong tidal torques considered here(the $\tau = 10$-year synchronization time-scale), tidal forces act to quickly restore equilibrium between the accretor (which becomes spun up at the onset of mass transfer) and the binary's orbit and the donor (which is spun down at the onset of mass transfer) and the orbit. Because the Roche-lobe radius is dependent upon the donor spin (see equation (\ref{scriptA})), the strength of tidal forces directly affects the mass-transfer rate.

Figure \ref{fig:evplot} shows the complete time-evolution of the chirp (top panel), mass transfer rate (middle panel), and gravitational wave frequency (bottom panel) for a system with initial donor and accretor masses of 0.25 $M_{\odot}$ and 0.55 $M_{\odot}$, respectively. The vertical solid green line marks the time of transition from direct-impact to disk accretion. As the top panel of Figure \ref{fig:evplot} shows, $\dot f_{\rm{astro}}$ dominates $\dot f_{\rm{GR}}$ over the course of the evolution, giving a negative $\dot f_{\rm{total}}$. The bottom panel shows the corresponding decrease in the gravitational wave frequency.

\begin{figure}
\plotone{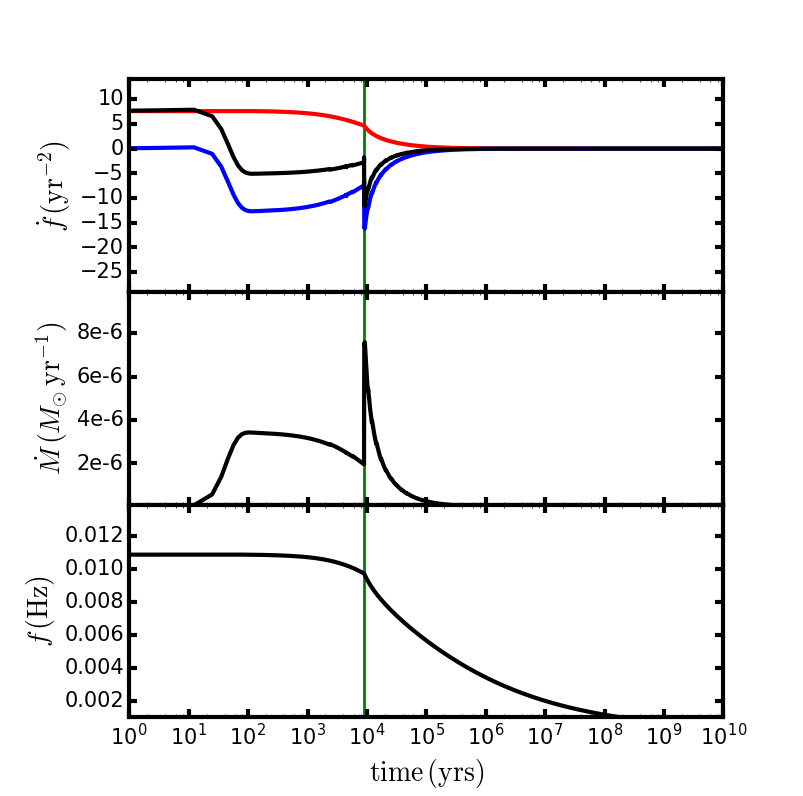}
\caption{\label{fig:evplot_225_1} \footnotesize Same as Figure \ref{fig:evplot}, but for a system with initial donor and accretor masses of 0.225 and 1.0 $M_{\odot}$, respectively. This system transitions from direct-impact to disk accretion after $8.92\times10^3$ years of evolution.}
\end{figure}

Figure \ref{fig:evplot_225_1} shows the long-term evolution of a system with initial donor and accretor masses of 0.225 and 1.0 $M_{\odot}$, respectively. As shown by Figure \ref{fig:plot}, this system remains stable throughout its entire evolution. As Figure \ref{fig:evplot_225_1} shows, the mass-transfer rate of this system experiences a brief spike upon the transition to disk accretion, and a corresponding spike in $\dot{f}_{\rm{MT}}$ and $\dot{f}_{\rm{total}}$. This spike in the mass-transfer rate is a consequence of the dependence of the Roche lobe upon the spin of the donor (equation \ref{scriptA}). As the donor leaves the direct-impact phase, it is slightly asynchronous with the orbit, giving a slightly higher value for the Roche lobe, and therefore a lower mass-transfer rate. Once the donor becomes fully synchronized upon the onset of disk accretion, the Roche lobe correspondingly shrinks, leading to the brief jump in $\dot{M}$. As seen in Figure \ref{fig:evplot}, the system with initial donor and accretor masses of 0.25 and 0.55 $M_{\odot}$ experiences this effect negligibly. This is explained by the fact that the 0.25-0.55 $M_{\odot}$ system remains in the direct-impact phase for a greater length of time than the 0.225-1.0 $M_{\odot}$ system ($3.79 \times 10^5$ years versus $8.92\times 10^3$ years) and therefore tidal forces have more time to bring the donor closer to full synchronization by the time disk accretion begins, leading to a negligibly small change in the Roche lobe and mass-transfer rate.

Figures \ref{fig:evplot} and \ref{fig:evplot_225_1} show that for both of these systems, the mass-transfer rate and $\dot{f}_{\rm{total}}$ taper to zero as the systems are driven to lower frequencies, throughout disk accretion, as expected.

\subsection{\textit{LISA} observations} \label{sec:LISA}

\subsubsection{Chirp}

\begin{figure*}
\begin{centering}
\includegraphics[width=0.7\textwidth]{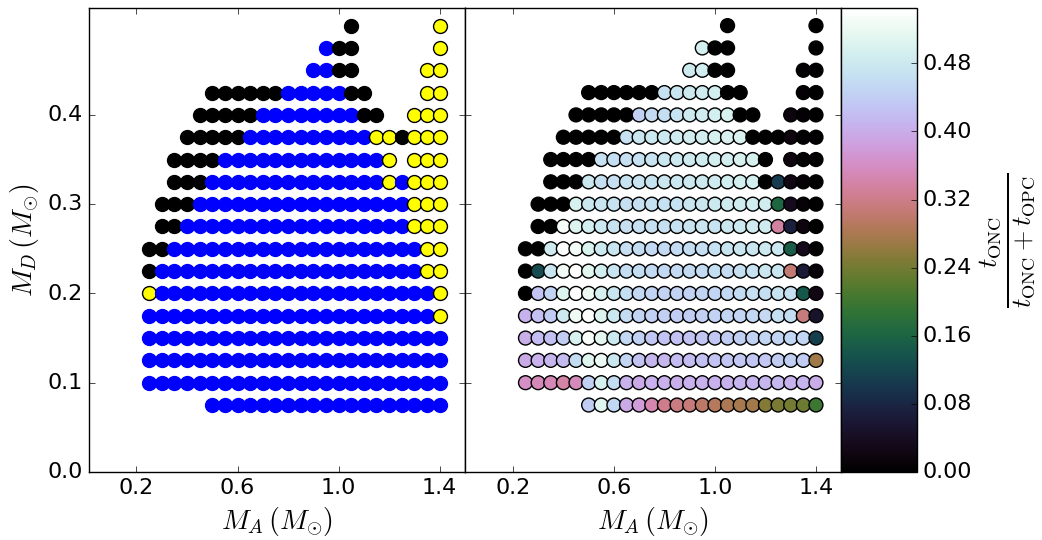}
\caption{\label{fig:timeplot} \footnotesize All systems from Figure \ref{fig:plot} that experience an observable negative chirp during their lifetime. In the left panel, the different colors denote the length of time, $t_{\rm{ONC}}$, each of these systems experiences an observable negative chirp. Systems marked with a blue circle have $t_{\rm{ONC}} >10^5$ years, yellow systems have $100 < t_{\rm{ONC}} < 10^5$ years, and systems marked with a black circle have $t_{\rm{ONC}} < 100$ years.  The right panel shows the fraction of the total observable lifetime each system exhibits an observable negative chirp, $t_{\rm{ONC}}/(t_{\rm{ONC}}+t_{\rm{OPC}})$.}
\end{centering}
\end{figure*}

As illustrated by Figures \ref{fig:chirp}, \ref{fig:evplot}, and \ref{fig:evplot_225_1}, it is possible for DWD systems to experience prominent negative chirps during their evolution. It will be possible for \textit{LISA} to measure the chirp of a detected binary with a chirp magnitude greater than $\sim0.1\ \rm{bin/yr}$  \citep{Takahashi2002} where the width of a frequency bin is determined by the length of the mission observation: 1 bin = $T_{\rm{obs}}^{-1}$. Following the recently accepted {LISA} mission proposal \citep{Amaro2017}, we take $T_{\rm{obs}}$ to be 4 years, giving us a minimum observable chirp of:
\begin{equation}
\dot{f}_{\rm{min}} = 0.1\,\rm{bin/yr} = 0.025\,\rm{yr}^{-2}  = 7.93\times 10^{-10}\,\rm{Hz}\,\rm{yr}^{-1}.
\end{equation}
Any chirp (positive or negative) with a magnitude greater than $\dot{f}_{\rm{min}}$ will be measurable by \textit{LISA}.

Figure \ref{fig:timeplot} shows the systems in our mass-mass parameter space (Figure \ref{fig:plot}) that experience a measurable negative chirp at some point during their evolution. In the left panel of the figure, the different colors denote the total length of time, $t_{\rm{ONC}}$, each of these systems experiences an observable negative chirp. Blue systems have $t_{\rm{ONC}} >10^5$ years, yellow systems have $100 < t_{\rm{ONC}} < 10^5$ years, and black systems have $t_{\rm{ONC}} < 100$ years. The blue systems clearly dominates the parameter space with 266 of the 450 total systems exhibiting negative chirps for at least $10^5$ years. The right panel of Figure \ref{fig:timeplot} shows the fraction of the total observable lifetime each system experiences an observable negative chirp, $t_{\rm{ONC}}/(t_{\rm{ONC}}+t_{\rm{OPC}})$. Here $t_{\rm{OPC}}$ is the total time each system exhibits an observable \textit{positive} chirp, which occurs during both inspiral when the binary is still detached and right at the onset of mass-transfer, just before the mass-transfer rate reaches its maximum value (see, for example, Figure 3).

As the right panel of Figure \ref{fig:timeplot} illustrates, 233 of the 450 total systems experience an observable negative chirp for $ \geq 40\%$ of their observable lifetime.

\subsubsection{Signal-to-noise ratio}

For circular DWD binaries, the scaling amplitude of the gravitational waves is given by:
\begin{equation}
\label{eq:h_o}
h_o = \frac{G}{c^2}\frac{M_c}{D}\left(\frac{G}{c^3}\pi f M_c \right)^{2/3},
\end{equation}
where $D$ is the distance of the binary from the detector and $M_c$ is the chirp mass, given by:

\begin{equation}
\label{chirp}
M_c = \frac{(M_A M_D)^{3/5}}{(M_A + M_D)^{1/5}}
\end{equation}
\begin{figure*} [t!] 
\begin{center}
\includegraphics[width=0.7\textwidth]{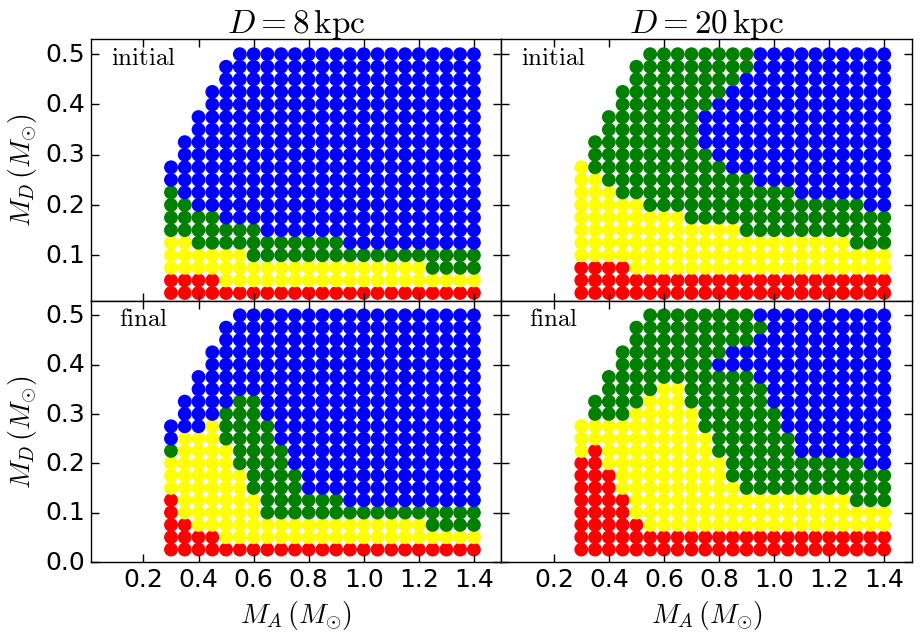}
\caption{\label{fig:SNRs} \footnotesize SNRs at the on-set of Roche-lobe overflow (top-panels) and at the conclusion of the direct-impact phase (bottom panels) for sources at distances of 8 kpc and 20 kpc.  Systems marked with blue circles have $\rm{SNR} > 10$, green have $5<\rm{SNR}<10$, yellow have $1<\rm{SNR}<5$, and red systems have $\rm{SNR}<1$.}
\end{center}
\end{figure*}

Here, as before, $T_{\rm{obs}} = 4$ years. The binaries considered in this analysis do not chirp appreciably over the course of a single observation. Therefore we can approximate the signal-to-noise ratio ($\rm{SNR}$) as:
\begin{equation}
\label{SNR}
\rm{SNR} \approx \frac{h_o \sqrt{T_{obs}}}{h_f},
\end{equation}
where $h_f$ is the spectral amplitude value for a specified gravitational-wave frequency, $f$, given by the standard \textit{LISA} sensitivity curve in \cite{Larson2002}.  For example, the maximal negative total chirp a DWD binary has in our simulations is $\approx -50$ $\rm{bins/yr}$ with a frequency of $f=0.03\ \rm{Hz}$ and the total chirp is not expected to evolve appreciably over the course of a \textit{LISA} observation. Thus approximating the total negative chirp as constant, the total frequency evolution of the binary is $\Delta f=1.59\times 10^{-6}\,\rm{Hz}$, which accounts for a total change in $\rm{SNR}$ of less than $0.01 \%$.

Figure \ref{fig:SNRs} shows the SNRs for the various initial masses at contact (top panels) and at the conclusion of the direct-impact phase (when the system becomes unstable or a disk is formed; bottom panels) for binaries at distances of 8 kpc, representing the Galactic center (left-hand panels) and 20 kpc, representing the Galactic edge (right-hand panels). Blue circles represent systems with $\rm{SNR}>10$, green systems with $5<\rm{SNR}<10$, yellow systems with $1<\rm{SNR}<5$, and red systems with $\rm{SNR}<1$. 

We expect all systems, with the exception of the red, to be resolvable by \textit{LISA} due to both their high frequency and high frequency evolution. At high frequencies, each DWD is expected to occupy its own frequency bin, and binaries with any measurable chirp are resolvable. Henceforth, we will define a resolvable binary as any binary with $\rm{SNR}>5$ and $|\,\dot{f}_{\rm{total}}| > 0.1\ \rm{bin/yr}$. 

As illustrated in Figure \ref{fig:timeplot}, $76\%$ (342 out of the available 450 systems) of the DWD systems in our parameter space experience a negative chirp prominent enough to be measured by \textit{LISA} for astrophysically relevant durations of time. Figure \ref{fig:SNRs} shows that, at the onset of mass-transfer,  
$80\%$ (362 out of 450) and $65\%$ (292 out of 450) of systems have an $\rm{SNR}>5$ for $D=8\,\rm{kpc}$ and $D=20\,\rm{kpc}$, respectively. We can make a more precise estimate on the number of systems that would have resolvable chirps and SNRs at the present day using Galactic population synthesis methods.

\section{Estimating the Milky Way DWD population}\label{sec:gxSimulation}

\begin{figure*}[t!] 
\begin{center}
\includegraphics[width=0.9\textwidth]{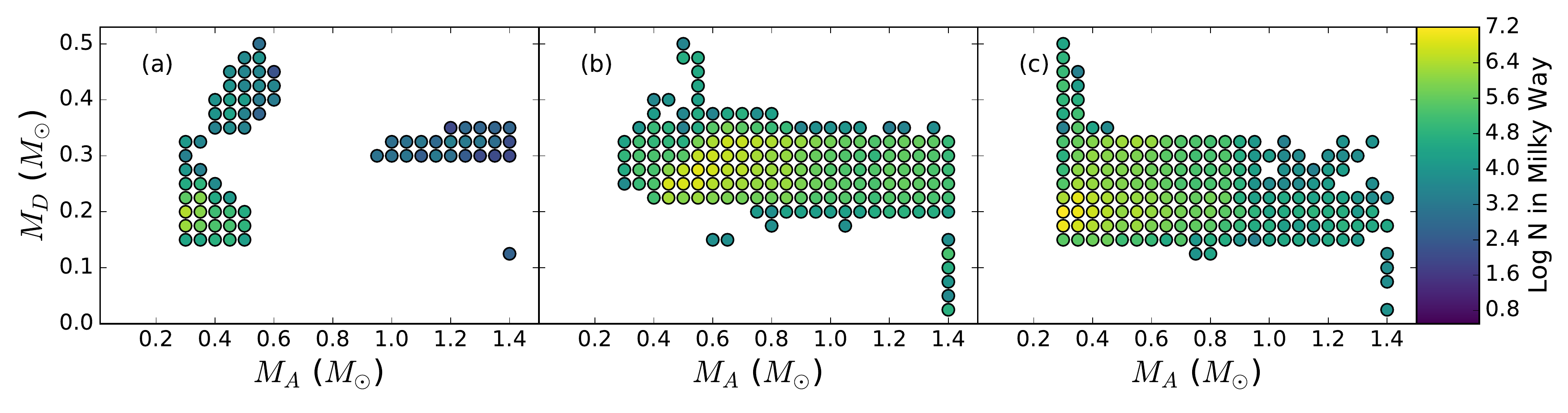}
\caption{\label{fig:MTonsetFixed} Mass distribution of He-donor DWDs that enter a mass transfer phase from both the Thin disk and Bulge. Panels (a), (b), and (c) show populations for the common envelope binding energy factor $\lambda=0.1,1.0,$ and $10.0$ respectively. The increase in number of systems as $\lambda$ increases is a result of the decreased binding energy of the donor envelope. This is also evident in the overall movement toward lower donor masses as $\lambda$ increases.}
\end{center}
\end{figure*}

To explore the number of He-donor DWD systems in the Milky Way, we generated population realizations using the binary stellar evolution code \texttt{BSE} \citep{Hurley2002}. \texttt{BSE} is a highly parallelizable open-source code that computes the evolution of a binary system from the birth of its stellar components to the final state of the binary, including potential compact remnants.

DWD progenitors are expected to pass through a common envelope (CE) phase en route to the formation of a DWD binary. The physical details of this CE phase (which are ripe with uncertainty) will have a substantial effect upon the orbital separations of the DWD binaries which emerge. To capture the uncertainty associated with the CE physics in our binary evolution models we simulated three separate DWD populations. Each population uses the $\alpha$-CE prescription detailed in \cite{Hurley2002} with $\alpha=1$ and a constant envelope binding energy factor $\lambda$, with the three models using $\lambda = 0.1, 1.0,$ and $10.0$.  All other stellar evolution prescriptions follow the fiducial model in \cite{Dominik2012}. Our choice in the range of $\lambda$ is motivated by analysis of \citet{Dewi2000}, which shows that for giant stars with masses less than ~$8\,M_{\odot}$ (which are consistent with WD progenitors at solar metallicity), $\lambda$ ranges from 0.1-10.\footnote{Note that the use of ion recombination energies to help eject the CE may be necessary to obtain high $\lambda$ values. \citep[e.g.,][]{Ivanova2013}.}

We simulate a population for both the Milky Way thin disk and the bulge.  We assume a constant star formation rate of $5\ M_{\odot} \rm{yr}^{-1}$ over $10\ \rm{Gyr}$ for the thin disk resulting in a total mass of $M_{\rm{thin\ disc}} = 5 \times 10^{10}\ M_{\odot}$ and a burst of star formation for the bulge 10 Gyr ago resulting in a total mass of $M_{\rm{bulge}} = 2\times10^{10} \ M_{\odot}$.

We assign initial binary states according to standard distribution functions as follows. Initial masses for the thin disk and bulge are distributed according to the \cite{Kroupa2001} initial mass function and $m^{-2.5}$ \citep{Robin2003}, respectively. Secondary masses are drawn from a uniform mass ratio \citep{Mazeh1992,Goldberg1994}. Initial separations are distributed flat in log-space at large separations (where $a \leq 5.75\times10^{6}R_{\odot}$) and fall off linearly at small separations as $(a/a_{0})^{1.2}$ for $a < 10R_{\odot}$ \citep{Han1998}. Finally, we use a thermal distribution for initial eccentricity \citep{Heggie1975}. 

The foundation of our simulations is a fixed population for each type of helium-donor DWD system: helium-helium (He-He), helium-carbon oxygen (He-CO), and helium-oxygen neon (He-ONe).  The fixed population is designed to be generated once and describe the physical parameters and relative rates of a given population that results from binary evolution.  Once a fixed population for a binary type (e.g. He-He) is simulated, we can Monte Carlo sample several hundred Milky Way realizations of the galactic population of that binary type.

The fixed population for a given type of binary must be large enough to encompass all possible physical parameters that could result from binary evolution.  To fulfill this requirement, we cumulatively simulate populations of binaries and use match criteria comparing  each successive simulation defined as:
\begin{equation} \label{eq: match}
 \rm{match} = \frac{\sum\limits_{k=1}^{N} P_{1} P_{2}}{ \sqrt{ \sum\limits_{k=1}^{N} (P_{1}P_{1})\sum\limits_{k=1}^{N} (P_{2}P_{2})}},
\end{equation}
\noindent where $P_{1}$ and $P_{2}$ are the histogram probabilities for bin $k$ in the successive simulations.  We simulate populations until the $\rm{match}$ exceeds $0.99$ or the change in $\rm{match}$ from the added population is less than $10^{-4}$ for each parameter of interest, in this case the donor and accretor masses and the time to mass-transfer onset. The match is highly sensitive to binwidth choice; here we use Knuth's Rule \citep{Knuth2006} implemented in Astropy \citep{Astropy}, a data-driven Bayesian binwidth selection selection method that enforces equal binwidths.

\begin{table}
\begin{center}
\caption{Density laws and scale heights.\label{tbl: gxdists}}
\begin{tabular}{ccc}
\tableline
\tableline
Galactic component & Density Law & Scale height \\
\tableline
Thin disk & $ \rho(R) \propto e^{-R/R_{0}}$ & $R_{0}=2.5$ kpc \\
 & $ \rho(z) \propto \rm{sech}^{2}(z/z_{0})$ & $z_{0}=0.352$ kpc \\
 Bulge & $ \rho(R) \propto  e^{-(R/R_{0})^2}$ & $R_{0}=0.5$ kpc \\
\tableline
\end{tabular}

\end{center}
\end{table}

We use \texttt{BSE} to evolve each population of DWD binaries up to the onset of mass transfer as determined by the Eggleton Roche-lobe-overflow criteria. Once DWD mass transfer begins in the \texttt{BSE} evolution algorithm, we log the mass of each component and the onset time of mass transfer, computed from the birth of both galactic components 10 Gyr ago. These are the parameters that make up the fixed population of DWDs at mass transfer onset, since all other binary parameters are either set by the mass transfer physics (e.g. separation) or are zero (e.g. eccentricity). Figure \ref{fig:MTonsetFixed} shows the distribution of donor and accretor masses for mass-transferring systems at the onset of mass-transfer for the Milky Way thin disk and bulge models. Comparing this figure to Figure \ref{fig:plot}, we note that DWD binaries with $M_{D}\lesssim0.15$ are generally not expected to undergo mass transfer in realistic galaxy models for all values of $\lambda$.

\begin{table*}[t!]
\begin{center}
\caption{Distribution statistics of the number of resolvable sources per realization for $100$ Milky Way realizations\label{tbl: stats}}
\begin{tabular}{cccccc}
\tableline
\tableline
Model, Chirp cut & $\rm{SNR}$ cut & $N_{min}$ & $N_{max}$ & $N_{ave}$ & Standard Deviation \\
\tableline
$\lambda=0.1,\ \dot{f}_{\rm{total}} < 0.0\ \rm{bin/yr}$ & $\rm{SNR} >1$ & 8931 & 9470 & 9235 & 100\\
  & $\rm{SNR} >5$ & 2862 & 3102 & 2991& 57\\
  & $\rm{SNR} >10$ & 1200 & 1394 & 1307 & 41\\
\tableline
$\lambda=1.0,\ \dot{f}_{\rm{total}} < 0.0\ \rm{bin/yr}$ & $\rm{SNR} >1$ & 88015 & 89171 &  88754 & 284\\
  & $\rm{SNR} >5$ & 24970 & 25795 & 25357 & 154\\
  & $\rm{SNR} >10$ & 10983 & 11552 & 11227 & 111\\
\tableline
$\lambda=10.0,\ \dot{f}_{\rm{total}} < 0.0\ \rm{bin/yr}$ & $\rm{SNR} >1$ & 122185 & 124067 &  123145 & 346\\
  & $\rm{SNR} >5$ & 38035 & 38991 & 38487 & 178\\
  & $\rm{SNR} >10$ & 16319 & 17012 & 16720 & 131\\

\tableline
\tableline
$\lambda=0.1,\ \dot{f}_{\rm{total}} < -0.1\ \rm{bin/yr}$ & $\rm{SNR} >1$ & 158 & 247 & 192 & 15\\
 & $\rm{SNR} >5$ & 154 & 241 & 187 & 15.4\\
 & $\rm{SNR} >10$ & 154 & 241 & 187 & 15.3\\
\tableline
$\lambda=1.0,\ \dot{f}_{\rm{total}} < -0.1\ \rm{bin/yr}$ & $\rm{SNR} >1$ & 2601 & 2854 & 2720 & 47\\
 & $\rm{SNR} >5$ & 2601 & 2854 & 2720 & 47\\
 & $\rm{SNR} >10$ & 2600 & 2853 & 2720 & 46.9\\
\tableline
$\lambda=10.0,\ \dot{f}_{\rm{total}} < -0.1\ \rm{bin/yr}$ & $\rm{SNR} >1$ & 2730 & 3995 & 2865 & 51.2\\
 & $\rm{SNR} >5$ & 2768 & 2934 & 2801 & 52\\
 & $\rm{SNR} >10$ & 2768 & 2931 & 2801 & 52\\
\tableline
\tableline
$\lambda=0.1,\ \dot{f}_{\rm{total}} < -1.0\ \rm{bin/yr}$ & $\rm{SNR} >1$ & 7 & 26 & 16 & 4.4\\
 & $\rm{SNR} >5$ & 7 & 26 & 16 & 4.4\\
 & $\rm{SNR} >10$ & 7 & 26 & 15 & 4.4\\
\tableline
$\lambda=1.0,\ \dot{f}_{\rm{total}} < -1.0\ \rm{bin/yr}$ & $\rm{SNR} >1$ & 277 & 349 & 314 & 14.6\\
 & $\rm{SNR} >5$ & 277 & 349 & 314 & 14.6\\
 & $\rm{SNR} >10$ & 277 & 349 & 314 & 14.6\\
\tableline
$\lambda=10.0,\ \dot{f}_{\rm{total}} < -1.0\ \rm{bin/yr}$ & $\rm{SNR} >1$ & 203 & 271 & 231 & 14\\
 & $\rm{SNR} >5$ & 198 & 266 & 228 & 13.9\\
 & $\rm{SNR} >10$ & 197 & 266 & 227 & 13.9\\
\tableline
\tableline
\tableline
\end{tabular}
\end{center}
\end{table*}

We then generate a realistic Milky Way population of DWDs at onset of mass transfer by Monte Carlo sampling the binary component masses and time of mass transfer from a three-dimensional Gaussian kernel density estimate. The thin disk and bulge populations are distributed with cylindrical and spherical symmetry respectively, with the distribution functions and scale heights defined in Table \ref{tbl: gxdists}\citep{Yu2011}.

Each DWD in the generated population is then evolved through phases of both direct-impact and/or disk mass transfer using the prescription outlined in section \ref{sec:calculation}. In this manner, we determine the binary parameters of each DWD at the present day, where the time to the present day is computed as
$t_{\rm{present}} = 10\ \rm{Gyr} - t_{\rm{MT-onset}}$. We retain only DWDs that are transferring mass at the present day.

\begin{figure}
\begin{center}
\includegraphics[width=0.48\textwidth]{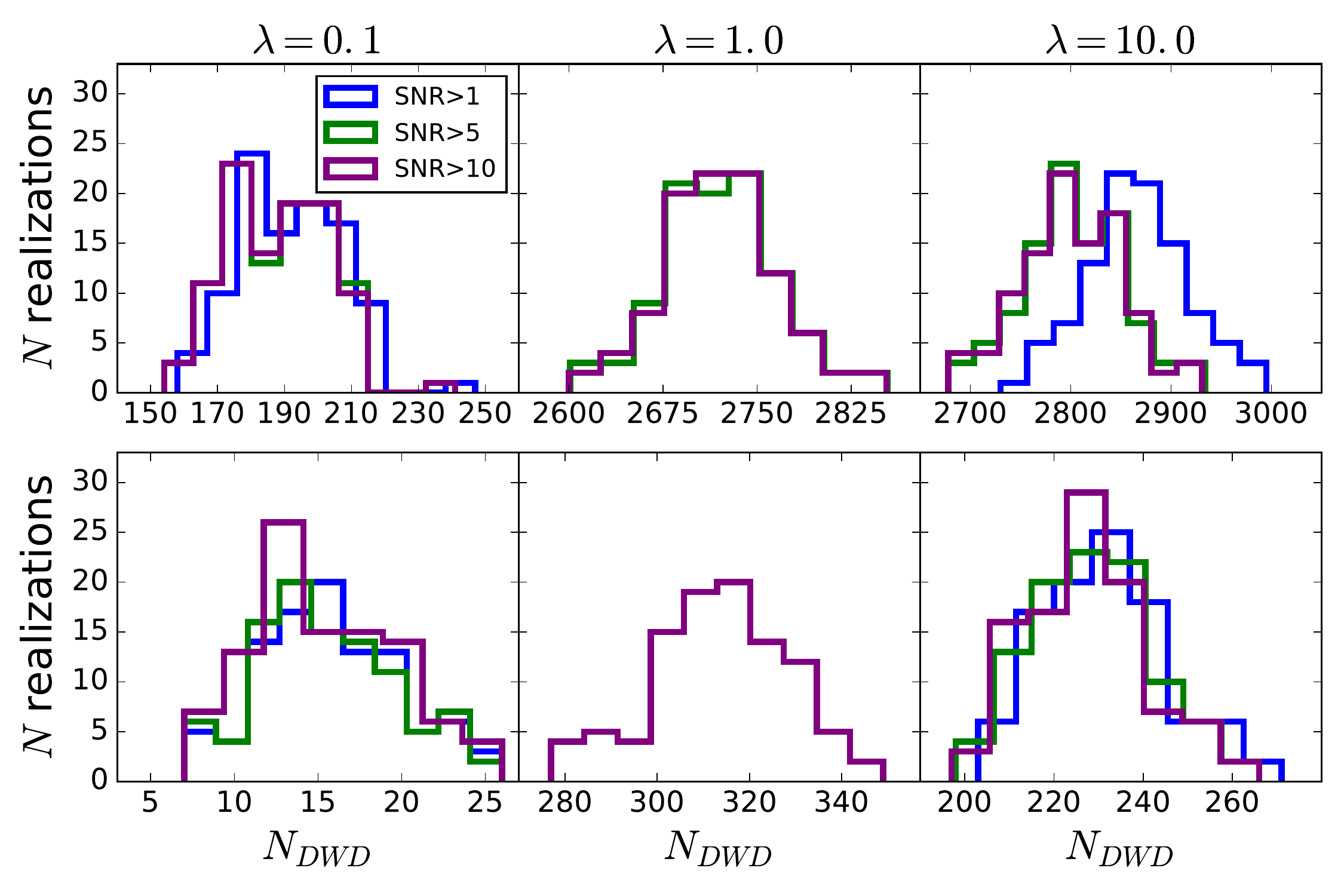}
\caption{\label{fig:gxHistos} Distributions of the number of sources with $\dot{f}_{\rm{total}}<-0.1\ \rm{bin/yr}$ (top row) and $\dot{f}_{\rm{total}}<-1.0\ \rm{bin/yr}$ (bottom row) for $100$ Monte-Carlo sampled Milky Way realizations for our $\lambda= 0.1$, (left column), $1.0$ (middle column), and $10.0$ (right column) models. We take three SNR cuts: $\rm{SNR}>1$,$5$, and $10$. Because the histograms converge for larger negative chirps, the three SNR plots overlap in the bottom row. See Table \ref{tbl: stats} for details.}
\end{center}
\end{figure}

We repeat this process to generate $100$ Milky Way realizations for each of our three common-envelope models. Figure \ref{fig:gxHistos} shows the distribution of the number of \textit{LISA}-resolvable sources with varying levels of negative total chirps for our $\lambda=0.1, 1.0$ and $10.0$ models.  For each model, the $\rm{SNR}$ distributions converge as the magnitude of the negative total chirp increases, leading to nearly identical distributions for $\dot{f}_{\rm{total}} < -1.0\ \rm{bin/yr}$. The distribution means and variances are shown in Table \ref{tbl: stats}.

For the rest of this paper, we define a fiducial SNR cut of $\rm{SNR}>5$ and consider two chirp cuts: $\dot{f}_{\rm{total}}< -0.1\ \rm{bin/yr}$ and $\dot{f}_{\rm{total}}< -1.0\ \rm{bin/yr}$.  In order to explore the population observable by \textit{LISA}, we select one Milky Way realization each from the $\lambda=0.1,\ 1.0$ and $10.0$ models. These realizations are selected based on the number of resolvable sources they contain, such that this number matches the average number of sources, $N_{ave}$, in Table \ref{tbl: stats}. For example, for the $\lambda=1.0$ model with $\rm{SNR}>5$, $N_{ave}\simeq2720$ and $N_{ave}\simeq313$ for $\dot{f}_{\rm{total}}<-0.1\ \rm{bin/yr}$ and $\dot{f}_{\rm{total}}<-1.0\ \rm{bin/yr}$ respectively. 

\begin{figure*}
\begin{center}
\includegraphics[width=0.7\textwidth]{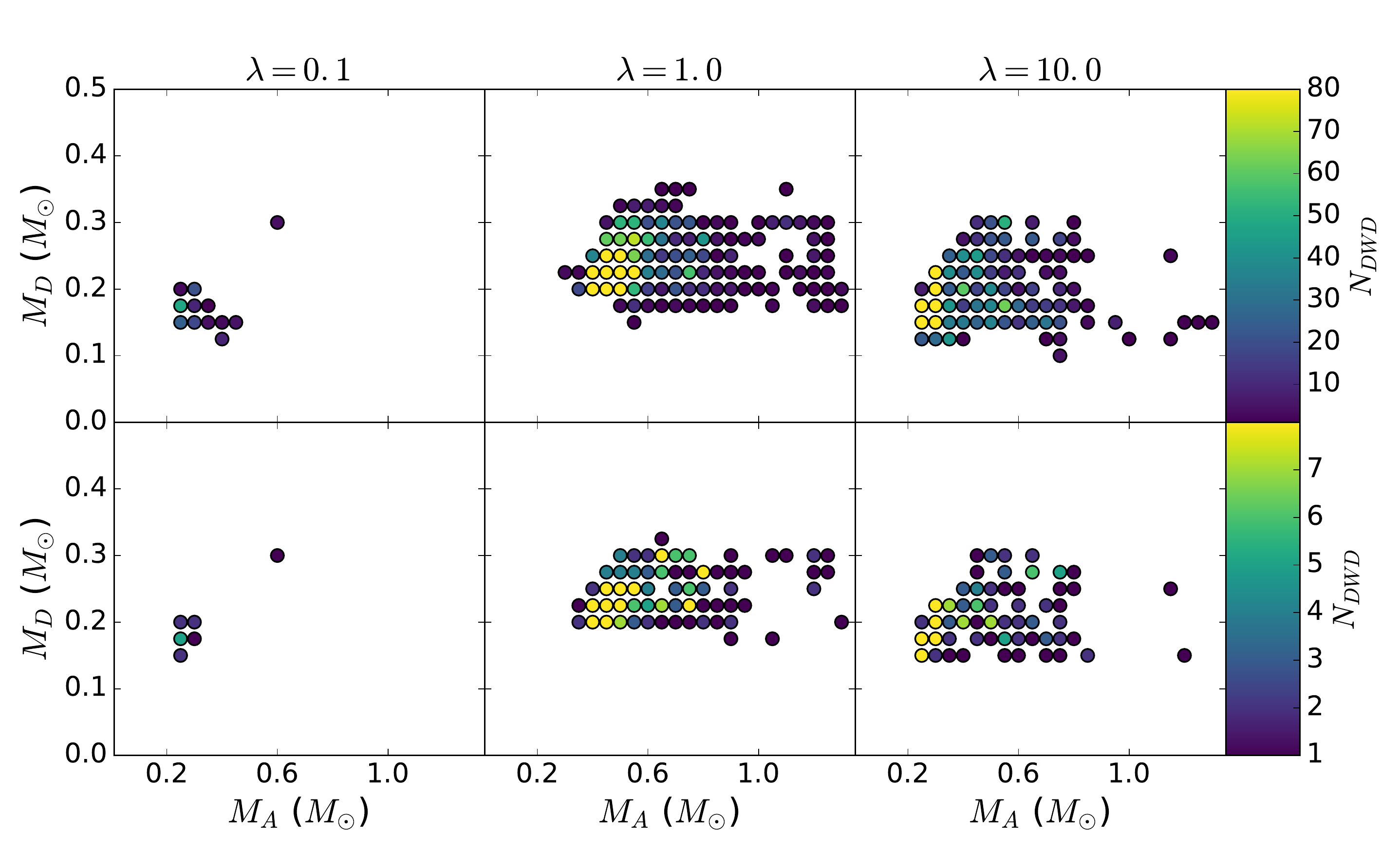}
\caption{\label{fig:gxExample}Mass distributions of galactic realization representative of the mean number of resolvable DWD binaries in their respective models. Each panel shows the donor and accretor masses at the onset of mass transfer for every binary with $\rm{SNR}>5$ and $\dot{f}_{\rm{total}}<-0.1\ \rm{bin/yr}$ (top row) or $\dot{f}_{\rm{total}}<-1.0\ \rm{bin/yr}$ (bottom row) at the present day. The color bar shows the number of systems that fall into each mass bin in the galactic realization.}
\end{center}
\end{figure*}

Figure \ref{fig:gxExample} shows the distribution of donor and accretor masses for these two realizations and our fiducial SNR and chirp cuts. The top (bottom) row shows the mass distribution for the subset of the galactic realization with $\rm{SNR}>5$ and $\dot{f}_{\rm{total}}<-0.1\ \rm{bin/yr}$ ($\dot{f}_{\rm{total}}<-1.0\ \rm{bin/yr}$), with many more systems fulfilling the less stringent chirp cuts. This is illustrated further in Figure \ref{fig:DistComparison} below.

As Figure \ref{fig:gxExample} illustrates, the majority of our potentially resolvable systems have components masses $\lesssim\,0.8\,M_{\odot}$,  which, for solar metallicity, point to progenitor masses $\lesssim 4\,M_{\odot}$. According to \citet{Dewi2000},  progenitors in this mass range point to $\lambda \sim 1$. 

\section{discussion} \label{sec:discussion}
 
\subsection{Observed DWD systems}
At present, there are no observed DWDs that exhibit prominent negative chirps as discussed in our analysis.

Two periodically varying sources with measured positive chirps have been discovered in recent years with periods of 10 minutes or less (RX J0806.3+1527, $P_{\rm{orb}}=321.5$ s and V407 Vul, $P_{\rm{orb}}=569.4$ s). The true nature of these systems is up for debate and three competing models have been proposed: (1) An intermediate polar model where the observed ultrashort periods are in not fact orbital, but spin periods of magnetic white dwarfs accreting from a non-degenerate companion (see, for example, \citet{Norton2004}), (2) a unipolar inductor model where the system is detached and the observed X-ray flux is electromagnetic in origin (see, for example, \citet{Wu2002,Dallosso2006,Dallosso2007}), and (3) an AM CVn model where the system is semi-detached and undergoing direct-impact accretion \citep[e.g.,][]{Marsh2002,Ramsay2002}.

The positive orbital frequency derivatives observed for both the RX J0806.3+1527 system, hereafter referred to as J0806, ($\dot{f}_{\rm{orb}} =3.77\pm 0.8\times10^{-16}$ Hz/s; \citet{Strohmayer2005}) and V407 Vul ($\dot{f}_{\rm{orb}} =3.77\pm 0.8\times10^{-16}$ Hz/s; \citet{Strohmayer2004a}) suggest the orbits of these two binaries are dominated by gravitational radiation. This suggests the AM CVn scenario is unlikely because stable mass transfer is expected to lead to a decrease in orbital frequency, as shown in this analysis as well as \citet{Kremer2015,Marsh2004,Gokhale2007}. However, alternative theories have been proposed in which a relatively low turn-on time-scale for mass transfer would allow for gravitational radiation to continue to dominate the orbital evolution even while the systems are accreting \citep[e.g.,][] {Deloye2006}.

As shown by Figures \ref{fig:evplot} and \ref{fig:evplot_225_1}, the calculations of this analysis suggest the time-scale for the mass-transfer to become the dominating factor in the orbital evolution is of order 100 years (defined as the time when $\dot{f}_{\rm{total}}$ becomes negative). This relatively short time-scale makes it unlikely we would catch a system in this phase of evolution. Therefore, on the basis of the results of our simulations, we believe the AM CVn model is not a viable explanation for these systems.

Instead, we favor the unipolar inductor model in which these observed systems are detached and gravitational radiation is currently driving the components together. We note that the observed orbital periods of these two systems are consistent with orbital periods of detached systems whose evolution is dominated by gravitational radiation.

In addition to J0806 and V407 Vul, a third system with an ultrashort period has been observed: ES Cet, $P_{\rm{orb}}=621$ s. Unlike J0806 and V407 Vul, observations of ES Cet suggest the presence of an accretion disk \citep{Warner2002,Strohmayer2004b}, though the true nature remains uncertain. 

A future LISA observation of  ES Cet could shed light on the details of its potential mass transfer, or lack thereof. Indeed, future observations of DWD systems in general by LISA will help to greatly constrain orbital parameters and binary evolution models, as discussed in section \ref{sec:CE}.

\subsection{Informing Common Envelope models}
\label{sec:CE}
\begin{figure}
\begin{center}
\includegraphics[width=0.4\textwidth]{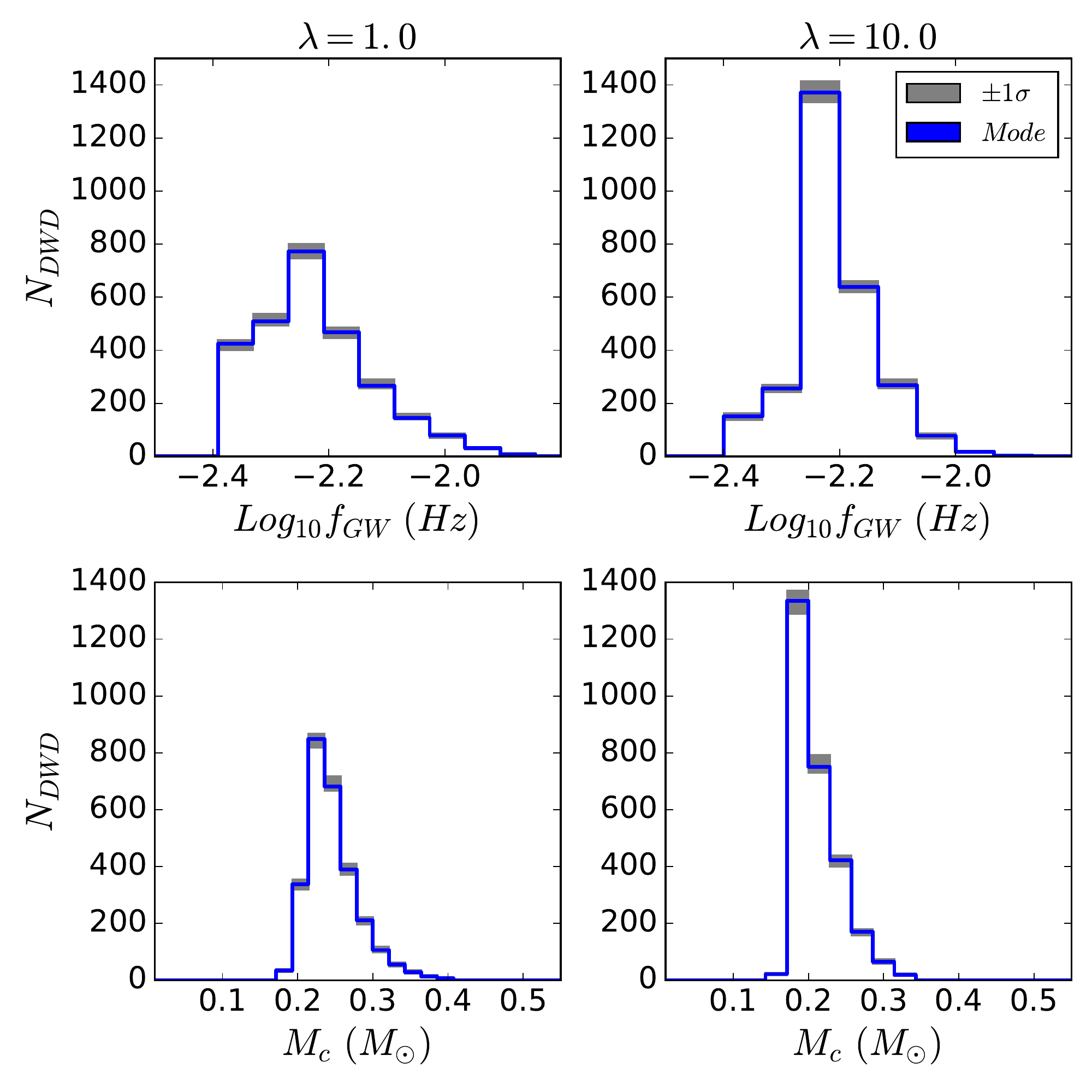}
\caption{\label{fig:DistComparison}Histograms of the gravitational-wave frequency,  $f_{\rm{GW}}$ and chirp mass, $M_c$, for DWD binaries with $\rm{SNR}>5$ and  $\dot{f}_{\rm{total}} < -0.1\ \rm{bin/yr}$. In blue are the histograms for $f_{\rm{GW}}$ and $M_c$ from the mode galactic realizations shown in Figure \ref{fig:gxExample}. The grey shaded regions show the $1\sigma$ deviations from the mode and the histogram bin widths are computed according to Knuth's Rule.}
\end{center}
\end{figure}

Figures \ref{fig:MTonsetFixed} and \ref{fig:gxExample} suggest that different common envelope binding energy prescriptions lead to different donor star and accretor star masses. Instead of measuring the donor and accretor masses separately, \textit{LISA} will measure the chirp mass of the system, defined as $M_c = (M_DM_A)^{3/5}/(M_D+M_A)^{1/5}$. Figure \ref{fig:DistComparison} shows the histograms of gravitational-wave frequency and chirp mass for the two example populations in Figure \ref{fig:DistComparison} with the same cuts considered. The histograms show a clear difference in both the frequency and chirp mass distributions, suggesting these systems as a way to inform common envelope prescriptions in binary evolution. 

To illustrate how the frequency and chirp mass distributions vary between galactic realizations, we plot histograms, where the bin height represents the mode bin height of our $100$ galactic realizations with the cutoffs  $\rm{SNR}>5$ and $\dot{f}_{\rm{total}}<-0.1\ \rm{bin/yr}$ for both the $\lambda=1.0$ and $10.0$ models in blue. We show in grey the $1\sigma$ deviations above and below the mode of each bin. The asymmetry in the deviations arises because we are using the mode instead of the mean of the distributions, which more accurately describes the range of heights for each bin. 

Here we neglect the $\lambda=0.1$ model due to the low number of expected chirping and resolvable systems leading to sparse histograms.  Furthermore, \citet{Dewi2000} suggests that $\lambda=0.1$ is unlikely for progenitors  with masses less than $10\,M_{\odot}$, with the majority of progenitors having values near $\lambda =1$.

While the bin heights of the histograms can vary considerably, the overall distribution shapes are consistent. This suggests that despite the cosmic variance that affects the LISA-observable population, different stellar envelope binding energy prescriptions lead to different observable populations, thus allowing a probe into the physics that determines the binding energy of stellar envelopes in the common envelope phase of binary evolution.
 
\subsection{Combining LISA and Gaia observations}

A parallax distance measurement, $D$, from \textit{Gaia} and gravitational-wave scaling amplitude, $h_o$, gravitational-wave frequency,  $f$ and total chirp measurement $\dot{f}_{\rm{total}}$ from \textit{LISA} allow the chirp mass to be computed directly (see equation \ref{eq:h_o}).  The chirp mass and gravitational-wave frequency completely determine the gravitational chirp, $\dot f_{\rm{GR}}$, which can be solved for from Eq. \ref{adot-GR}. Then $\dot f_{\rm{GR}}$ and $\dot{f}_{\rm{total}}$ can be combined to solve for the overall astrophysical chirp:
\begin{equation}
\dot{f}_{\rm{astro}} = \dot{f}_{\rm{total}} - \dot{f}_{\rm{GR}},  
\end{equation}
where $\dot f_{\rm{astro}} = \dot f_{\rm{MT}} + \dot f_{\rm{tides}}$.  A deeper study of the resolvability of the mass-transferring DWDs of the Milky Way by both \textit{Gaia} and \textit{LISA} is currently in progress \citep{Breivik2017}.

\section{Conclusion} \label{sec:conclusion}
We have expanded upon the earlier analysis of \citet{Kremer2015} to explore the long-term evolution of mass-transferring DWD systems, which includes treatment of these systems through phases of both direct-impact and disk accretion. Additionally, we have self-consistently modeled these systems through phases of super-Eddington accretion.

By examining the chirps and SNRs of these systems, we have shown that a substantial fraction of mass-transferring DWD binaries (both direct-impact and disk accretors) will be observable by \textit{LISA} and that these systems exhibit prominent negative chirps as a consequence of the dominant role of mass transfer in their long-term evolution.

Additionally, we generated 100 galactic population realizations to estimate the number of mass-transferring DWDs in the Milky Way that will observable by {LISA}. We predict $2720$ DWDs observable with  $\rm{SNR} > 5$ and $\dot{f}_{\rm{total}} < -0.1\ \rm{bin/yr}$, and 227 DWDs with
 $\rm{SNR}>10$ and  $\dot{f}_{\rm{total}} < -1.0\ \rm{bin/yr}$. The majority of these sources have exceptionally large SNRs due to their nearby location in the Milky Way, suggesting that a complementary observation by \textit{Gaia} may be possible. These systems, observed both by \textit{LISA} and \textit{Gaia}, will allow an exquisite characterization of DWD mass-transferring systems.

%%%%%%%%%%%%%%%%%%%%%%%%%%%%%%%%%3

\acknowledgments

We thank Jeff Andrews for useful discussions and the referee for useful suggestions and inquires.

KK acknowledges support from the National Science Foundation Graduate Research Fellowship Program under Grant No. DGE-1324585. KB and SLL acknowledge support from NASA Grant NNX13AM10G. VK acknowledges support from Northwestern University. The majority of our analysis was performed using the computational resources of the Quest high performance computing facility at Northwestern University which is jointly supported by the Office of the Provost, the Office for Research, and Northwestern University Information Technology.

\software{BSE \citep{Hurley2002}}

\listofchanges


\vspace{5mm}


\begin{thebibliography}{}
\bibitem [Althaus et al.\,(2013)] {Althaus2013} Althaus, L. G., Miller Bertolami, M. M., \& Córsico, A. H. 2013, A\&A, 557, A19
\bibitem[Amaro-Seoane et al.\,(2013)] {Amaro2013} Amaro-Seoane, P. et al. 2013, arXiv: 1305.5720
\bibitem[Amaro-Seoane et al.\,(2017)] {Amaro2017} Amaro-Seoane, P. et al. 2017, arXiv: 1702.00786
\bibitem[Astropy Collaboration et al.\,(2013)]{Astropy} Astropy Collaboration, Robitaille, T.~P., Tollerud, E.~J., et al.\ 2013, A\&A, 558, A33 
\bibitem[Breivik et al.$\,$(2017)] {Breivik2017} Brevik, K. et al. 2017, in progress
\bibitem [Dall'Osso et al.$\,$(2006)] {Dallosso2006} Dall'Osso, S., Israel, G.L., and Stella, L. 2006, A\&A, 447, 785
\bibitem [Dall'Osso et al.$\,$(2007)] {Dallosso2007} Dall'Osso, S., Israel, G.L., and Stella, L. 2007, A\&A, 464, 417
\bibitem[Deloye \& Taam$\,$(2006)] {Deloye2006} Deloye, C.J. and Taam, R.E. 2006, ApJ, 649, L99
\bibitem[Dewi \& Tauris(2000)] {Dewi2000} Dewi,  J.D.M. and Tauris, T.M. 2000, A\&A 360,1043
\bibitem[Dominik et al.(2012)]{Dominik2012} Dominik, M., Belczynski, K., Fryer, C., et al.\ 2012, \apj, 759, 52 
\bibitem[Eggleton$\,$(1983)] {Eggleton1983} Eggleton, P. P. 1983, ApJ, 268, 368.
\bibitem[Frank, King and Raine$\,$(2002)] {Frank2002} Frank, J., King, A., and Raine, D. 2002, Accretion Power in Astrophysics. Cambridge University Press, Cambridge.
\bibitem[Fuller \& Lai$\,$(2014)] {Fuller2014} Fuller, J., and Lai, D. 2014, MNRAS, 444, 3488.
\bibitem [Gokhale et al.$\,$(2007)] {Gokhale2007} Gokhale, V., Peng, X. M. and Frank, J. 2007, ApJ 655, 1010.
\bibitem[Goldberg \& Mazeh(1994)]{Goldberg1994} Goldberg, D., \& Mazeh, T.\ 1994, \aap, 282, 801 
\bibitem[Han(1998)]{Han1998} Han, Z.\ 1998, \mnras, 296, 1019 
\bibitem [Han and Webbink$\,$(1999)] {Han1999} Han, Z. and Webbink, R. F. 1999 A and A, 349, L17.
\bibitem[Hurley et al.$\,$(2002)]{Hurley2002} Hurley, J. R., Tout, C. A., \& Pols, O. R. 2002, MNRAS, 329, 897.
\bibitem[Heggie(1975)]{Heggie1975} Heggie, D.~C.\ 1975, \mnras, 173, 729 
\bibitem[Iben \& Tutukov\,(1984)]{Iben1984} Iben,I. \& Tutukov, A.V 1984, ApJ 54, 335
\bibitem[Ivanova et al.$\,$(2013)] {Ivanova2013} Ivanova, N., Justham, S., Chen, X., et al. 2013, A\&ARv, 21, 59
\bibitem[Knuth(2006)]{Knuth2006} Knuth, K.~H.\ 2006, arXiv:physics/0605197 
\bibitem[Kremer et al.(2015)]{Kremer2015} Kremer, K., Sepinsky, J., \& Kalogera, V. \ 2015, \apj, 806, 76.
\bibitem[Kroupa(2001)]{Kroupa2001} Kroupa, P.\ 2001, \mnras, 322, 231 
\bibitem[Larson et al.(2002)]{Larson2002} Larson, S.~L., Wellings, R.~W., \&Hiscock, W.~A., \prd, 66, 062001 
\bibitem[Liu et al.(2010)]{Liu2010} Liu, J., Han, Z., Zhang, F., \& Zhang, Y.\ 2010, \apj, 719, 1546  
\bibitem[Marsh et al.$\,$(1995)] {Marsh1995} Marsh, T. R., Dhillon, V. S. and Duck, S. R. 1995, MNRAS, 275, 828.
\bibitem[Marsh \& Steeghs$\,$(2002)] {Marsh2002} Marsh, T.R. and Steeghs, D. 2002, \mnras, 331, L7
\bibitem [Marsh et al.$\,$(2004)] {Marsh2004} Marsh, T. R., Nelemans, G. and Steeghs, D. 2004, MNRAS, 350, 113.
\bibitem[Marsh(2011)]{Marsh2011} Marsh, T.~R.\ 2011, Classical and Quantum Gravity, 28, 094019 
\bibitem [Maoz et al.$\,$(2014)] {Maoz2014} Maoz, D., Mannucci, F., and Nelemans, G. 2014, AARA, 52, 107.
\bibitem[Mazeh et al.(1992)]{Mazeh1992} Mazeh, T., Goldberg, D., Duquennoy, A., \& Mayor, M.\ 1992, \apj, 401, 265 
\bibitem [Motl et al.\,(2007)] {Motl2007} Motl, P. M., Frank, J., Tohline, J. E., \& D’Souza, M. C. R. 2007, ApJ, 670, 1314
\bibitem [Nather et al.$\,$(1981)] {Nather1981} Nather, R. E., Robinson, E. L. and Stover, R. J. 1981, ApJ, 244, 269.
\bibitem[Nelemans et al.(2001a)]{Nelemans2001a} Nelemans, G., Yungelson, L.~R., Portegies Zwart, S.~F., \& Verbunt, F.\ 2001, \aap, 365, 491
\bibitem[Nelemans et al.(2001b)]{Nelemans2001b} Nelemans, G., Portegies Zwart, S.~F., Verbunt, F., \& Yungelson, L.~R.\ 2001, \aap, 368, 939
\bibitem[Nelemans et al.(2001c)]{Nelemans2001c} Nelemans, G., Yungelson, L.~R., \& Portegies Zwart, S.~F.\ 2001, \aap, 375, 890 
\bibitem[Nelemans et al.(2004)]{Nelemans2004} Nelemans, G., Yungelson, L.~R., \& Portegies Zwart, S.~F.\ 2004, \mnras, 349, 181 
\bibitem[Norton et al.$\,$(2004)] {Norton2004} Norton, A.J., Haswell, C.A. and Wynn, G.A. 2004, A\&A 419, 1025.
\bibitem [Ramsay et al.$\,$(2002)] {Ramsay2002} Ramsay, G., Wu, K., Cropper, M., Schmidt, G., Sekiguchi, K. Iwamuro, F., and Maihara, T. 2002, \mnras, 333, 575
\bibitem [Ramsay et al.$\,$(2005)] {Ramsay2005} Ramsay, G., Hakala, P., Wu, K. et al. 2005, MNRAS 357, 49.
\bibitem[Robin et al.(2003)]{Robin2003} Robin, A.~C., Reyl{\'e}, C., Derri{\`e}re, S., \& Picaud, S.\ 2003, \aap, 409, 523 
\bibitem[Ruiter et al.(2010)]{Ruiter2010} Ruiter, A.~J., Belczynski, K., Benacquista, M., Larson, S.~L., \& Williams, G.\ 2010, \apj, 717, 1006 
\bibitem[Sepinsky \& Kalogera(2014)]{Sepinsky2014} Sepinsky, J. \& Kalogera, V. \ 2014, \apj, 785, 157
\bibitem[Sepinsky et al.$\,$(2007)] {Sepinsky2007} Sepinsky, J., Willems, B. and Kalogera, V. 2007, ApJ, 660, 1624.
\bibitem[Shen$\,$(2015)] {Shen2015} Shen, K. 2015, ApJL,805, L6.
\bibitem[Soberman et al.$\,$(1997)] {Soberman1997} Soberman, G.E., Phinney, G.S., and van den Huevel, E. P. J. 1997, A and A, 327, 620.
\bibitem[Strohmayer$\,$(2004a)] {Strohmayer2004a} Strohmayer, T. 2004a, ApJ 610, 416.
\bibitem[Strohmayer$\,$(2004b)] {Strohmayer2004b} Strohmayer, T. 2004b, ApJ 614, 358.
\bibitem[Strohmayer$\,$(2005)] {Strohmayer2005} Strohmayer, T. 2005, ApJ, 627, 920
\bibitem[Takahashi \& Seto(2002)]{Takahashi2002} Takahashi, R., \& Seto, N.\ 2002, \apj, 575, 1030 
\bibitem[Tutukov and Yungelson$\,$(1996)] {Tutukov1996} Tutukov, A. and Yungelson, L. 1996, MNRAS, 280, 1035
\bibitem [Verbunt and Rappaport$\,$(1988)] {Verbunt1988} Verbunt, F. and Rappaport, S. 1988, ApJ, 332, 193
\bibitem [Warner and Woudt$\,$(2002)] {Warner2002} Warner, B. and Woudt, P. 2002, PASP 792, 129
\bibitem [Webbink\,(1984)] {Webbink1984} Webbink, R.F. 1984, ApJ 277, 355
\bibitem [Wu et al.$\,$(2002)] {Wu2002} Wu, K., Cropper, M., Ramsay, G. and Sekiguchi, K. 2002, \mnras, 331, 221
\bibitem[Yu \& Jeffery(2010)]{Yu2010} Yu, S., \& Jeffery, C.~S.\ 2010, \aap 521, A85 
\bibitem[Yu \& Jeffery(2011)]{Yu2011} Yu, S., \& Jeffery, C.~S.\ 2011, \mnras,417, 1392 
\bibitem[Zahn\,(1977)]{Zahn1977} Zahn, J.-P. 1977, A\&A 57, 383


\end{thebibliography}
\end{document}